\documentclass{article} \usepackage{slashed,feynmf,epsfig,amsmath,amssymb,enumitem} \usepackage[papersize={8.5in,11in}]{geometry}
  \renewcommand{\vec}[1]{\boldsymbol{\mathrm{#1}}} \usepackage{bm}
\usepackage{graphicx}
\begin{document}
\title{Acceleration in Modified Gravity (MOG) and the Mass-Discrepancy Baryonic Relation}
\author{J. W. Moffat\\
Perimeter Institute for Theoretical Physics, Waterloo, Ontario N2L 2Y5, Canada\\
and\\
Department of Physics and Astronomy, University of Waterloo, Waterloo,\\
Ontario N2L 3G1, Canada}
\maketitle



\begin{abstract}The equation of motion in the generally covariant modified gravity (MOG) theory leads for weak gravitational fields and the non-relativistic limit to a modification of the Newtonian gravitational acceleration law, expressed in terms of two parameters $\alpha$ and $\mu$. The parameter $\alpha$ determines the strength of the gravitational field and $\mu$ is the effective mass of the vector field $\phi_\mu$, coupled with gravitational strength to baryonic matter. The MOG acceleration law for weak field gravitation and non-relativistic particles has been demonstrated to fit a wide range of galaxies, galaxy clusters and the Bullet Cluster and Train Wreck Cluster mergers. We demonstrate that the MOG acceleration law for a point mass source is in agreement with the McGaugh et al., correlation between the radial acceleration traced by galaxy rotation curves and the distribution of baryonic matter for the SPARC sample of 153 rotationally supported spiral and irregular galaxies.
\end{abstract}

\maketitle


\section{Introduction}

The pioneering research by Zwicky~\cite{Zwicky} and Vera Rubin and collaborators~\cite{Rubin1,Rubin2} showed that the dynamics of galaxies and galaxy clusters did not agree with the predictions of Newtonian and Einstein gravity. The masses of galaxies and clusters inferred from dynamics were found to exceed the baryon mass in these systems. This fact promoted the need for a mysterious dark matter to form halos around galaxies, resolving the discrepancy with gravity. To date, there is no convincing evidence either in deep underground laboratory experiments such as LUX~\cite{LUX} and Panda-X~\cite{PandaX}, and astrophysical and LHC experiments to support the existence of exotic dark matter particles. Modified gravitation is an alternative explanation for the dynamics of galaxies and clusters. A fully covariant and relativistic theory of gravitation called Scalar-Tensor-Vector-Gravity (STVG) or MOG (modified gravity) has been developed to explain the dynamics of galaxies, clusters and the large-scale structure of the universe without detectable dark matter~\cite{Moffat1}.

We require that our gravitational theory fulfills the following:

\begin{enumerate}

\item General covariance and local Lorentz invariance,

\item Classical causal locality,~\footnote[1]{The requirement (2) may not be true for a quantum gravity theory for which the requirement of ultraviolet completeness may demand violation of locality at small distances of order the Planck length, ${\ell}_{\rm Pl}=(\hbar G_N/c^3)=1.62\times 10^{-35}\,{\rm m}$~\cite{Moffat2}.}

\item Equivalence principle,

\item Contains General Relativity (GR) in a well-defined limit.

\end{enumerate}

We have shown that MOG fits a large number of galaxy rotation curves~\cite{MoffatRahvar1}, satisfies the Tully-Fisher relation~\cite{TullyFisher,Verheijen} and also describes successfully the dynamics of clusters~\cite{MoffatRahvar2} and merging clusters, such as the Bullet Cluster and the Train Wreck Cluster Abell 520~\cite{BrownsteinMoffat,IsraelMoffat}. In the following, we will demonstrate the consistency of MOG with an empirical relation between acceleration and baryonic matter for galaxies determined by McGaugh et al.,~\cite{McGaugh}.

\section{MOG Acceleration Law}

The equation of motion for a massive test particle in MOG has the covariant form~\cite{Moffat1}:
\begin{equation}
\label{eqMotion}
\frac{du^\mu}{ds}+{\Gamma^\mu}_{\alpha\beta}u^\alpha u^\beta=\frac{q_g}{m}{B^\mu}_\nu u^\nu,
\end{equation}
where $u^\mu=dx^\mu/ds$ with $s$ the proper time along the particle trajectory, ${\Gamma^\mu}_{\alpha\beta}$ denote the Christoffel symbols, $B_{\mu\nu}=\partial_\mu\phi_\nu-\partial_\nu\phi_\mu$ and $\phi_\mu$ is a massive vector field. Moreover, $q_g$ and $m$ denote the MOG gravitational test particle charge $q_g=\sqrt{\alpha G_N}m$ and the test particle mass $m$, respectively, and $G_N$ is Newton's gravitational constant. In the following, we define the gravitational interaction strength to be $G=G_N(1+\alpha)$. The equation of motion for a photon is~\footnote[2]{The null geodesic for a photon is screened by a conformal metric~\cite{Moffat3}.} 
\begin{equation}
\frac{du^\mu}{ds}+{\Gamma^\mu}_{\alpha\beta}u^\alpha u^\beta=0.
\end{equation}
We note that for $q_g/m=\sqrt{\alpha G_N}$ the equation of motion for a test particle (\ref{eqMotion}) {\it satisfies the (weak) equivalence principle}, leading to the free fall of particles in a homogeneous gravitational field, although the free-falling particles do not follow geodesic motion. MOG contains GR in the limit that the scalar field parameter $\alpha=0$. 

In the weak field slow motion approximation, $dr/ds\sim dr/dt$ and $2GM/r\ll 1$ the radial motion acceleration of a test particle is given by
\begin{equation}
\frac{d^2r}{dt^2}+\frac{GM}{r^2}=\frac{q_gQ_g}{m}\frac{\exp(-\mu r)}{r^2}(1+\mu r),
\end{equation}
where $Q_g=\sqrt{\alpha G_N}M$ with $M$ equal to the source mass. For $ q_gQ_g/m=\alpha G_NM$ and $G=G_N(1+\alpha)$, the modified Newtonian acceleration law for a point particle is~\cite{Moffat1}:
\begin{equation}
\label{MOGaccelerationlaw}
a_{\rm MOG}(r)=-\frac{G_NM}{r^2}\biggl[1+\alpha-\alpha\exp(-\mu r)\biggl(1+\mu r\biggr)\biggr].
\end{equation}
The acceleration law can be extended to a distribution of matter:
\begin{equation}
\label{accelerationlaw2}
a_{\rm MOG}({\vec x})=-G_N\int d^3{\vec x}'\frac{\rho_{\rm bar}({\vec x}')({\vec x}-{\vec x}')}{|{\vec x}-{\vec x}'|^3}
[1+\alpha-\alpha\exp(-\mu|{\vec x}-{\vec x}'|(1+\mu|{\vec x}-{\vec x}'|)],
\end{equation}
where $\rho_{\rm bary}(\bf x)$ is the total baryon density of matter.

In the weak gravitational field limit, the MOG field equations (see Appendix A) lead to the point particle potential for a static spherically symmetric system:
\begin{equation}
\Phi(r)=\Phi_N(r)+\Phi_Y(r),
\end{equation}
where
\begin{equation}
\Phi_N(r)=-\frac{G_{\infty}M}{r}=-\frac{G_N(1+\alpha)M}{r},
\end{equation}
and
\begin{equation}
\Phi_Y(r)=\frac{\alpha G_NM\exp(-\mu r)}{r}.
\end{equation}
For continuous distributions of matter with baryon density $\rho_{\rm bar}$, the $\Phi_N$ and $\Phi_Y$ potentials are given by
\begin{equation}
\Phi_N({\vec x})=-G_N(1+\alpha)\int d^3{\vec x}'\frac{\rho_{\rm bar}({\vec x}')}{|{\vec x}-{\vec x}'|},
\end{equation}
and
\begin{equation}
\Phi_Y({\vec x})=\alpha G_N\int d^3{\vec x}'\frac{\exp(-\mu|{\vec x}-{\vec x}'|)\rho_{\rm bar}({\vec x}')}{|{\vec x}-{\vec x}'|}.
\end{equation}
The complete MOG potential for a given density $\rho_{\rm bar}({\vec x})$ is
\begin{equation}
\Phi(\vec{x}) = - G_N \int d^3x'\frac{\rho_{\rm bar}(\vec{x}')}{|\vec{x}-\vec{x}'|}\Big[1+\alpha -\alpha \exp(-\mu|\vec{x}-\vec{x}'|)\Big].
\label{potential}
\end{equation}
For a delta-function source density $\rho_{\rm bar}({\vec x})=M\delta^3({\vec x})$.

The Poisson equations for $\Phi_N(r)$ and $\Phi_Y(r)$ are given by
\begin{equation}
\label{NewtonPot}
\nabla^2\Phi_N(r)=4\pi G\rho_{\rm bar}(r),
\end{equation}
and
\label{YukawaPot}
\begin{equation}
(\nabla^2-\mu^2)\Phi_Y(r)=-4\pi\alpha G_N\rho_{\rm bar}(r),
\end{equation}
respectively.

MOG can be considered a fundamental classical theory of gravitation, for it is based on a fully covariant action principle and field equations. It reduces to GR in the well-defined limit $\alpha=\alpha(x)=0$. In contrast, MOND~\cite{Milgrom1,Milgrom2} is an empirical non-relativistic formula which by itself is not based on a covariant fundamental action principle. It provides a fit to galaxy accelerations and velocity rotation curves with a characteristic acceleration scale $a_0=1.2\times 10^{-10}\,{\rm m/s^2}$ by empirically choosing from many possible interpolating functions but it fails to describe clusters and cosmology without dark matter.
 
\section{MOG Flat Rotation Curves}

Observations of galaxies show a flat rotation curve at the edge of galaxies with visible stellar mass and gas. This is not consistent with Newtonian gravity. The MOG acceleration law has been demonstrated to agree with observational data for galaxies and clusters. To obtain a qualitative understanding of the dynamics and flat rotation curves of galaxies in MOG, we will restrict ourselves to the spherically symmetric point particle solution. A more rigorous determination of rotational velocity curves for galaxies requires using the continuous distribution of matter~\cite{MoffatRahvar1}.  

In the limit that $r\rightarrow\infty$, we get from (\ref{MOGaccelerationlaw}) for approximately constant $\alpha$ and $\mu$:
\begin{equation}
\label{AsymptoticMOG}
a(r)\approx -\frac{G_N(1+\alpha)M_{\rm bar}}{r^2}.
\end{equation}
MOG has a Newtonian-Kepler behavior for large $r$ with enhanced gravitational strength $G=G_N(1+\alpha)$. The transition from Newtonian acceleration behavior for small $r$ to non-Newtonian behavior for intermediate values of $r$ is due to the repulsive Yukawa contribution in (\ref{MOGaccelerationlaw}). This can also result in the rotational velocity having a maximum value in the transition region.

To obtain the maximum, horizontally flat rotational velocity $V_{\rm flat}(R_{\rm flat})$, we differentiate the rotational velocity $v_c(r)$, $v'_c(r)=dv_c(r)/dr$ and set it equal to zero. For circular velocities, we have $v_c(r)=\sqrt{-a(r)r}$ which gives
\begin{equation}
v'_c(r)=-\frac{a(r)+ra'(r)}{2\sqrt{-a(r)r}}.
\end{equation}
Thus, $v_c'(r)$ vanishes if $a'(r)=-a(r)/r$. We obtain from (\ref{MOGaccelerationlaw}):
\begin{equation}
\label{MaxMOGvel}
V_{\rm flat MOG} = (G_N M_{\rm bar} \alpha R_{\rm flat})^{1/2}\mu\exp(-\mu R_{\rm flat}/2).
\end{equation}
The vanishing of $v'_c$ occurs for $\mu=1/R_{\rm flat}$ and $\alpha=e/(3-e)=9.64894$ and is the condition for the rotational velocity curve to become horizontal. If $\alpha$ is less than $\alpha=9.64894$ the rotation curves will be flattened but not exactly horizontally flat, while for greater values of $\alpha=9.64894$, the rotational velocity curves have a region of rising slope, which would be a region in a galaxy with no stable circular orbits. The value $\alpha=9.64894$ is close to the best fit value $\alpha=8.89\pm 0.34$ for fitting MOG to rotational velocity data~\cite{MoffatRahvar1}. We obtain for circular velocities from (\ref{MaxMOGvel}):
\begin{equation}
\label{Maxaccel}
a_{{\rm flat\,MOG}} = -G_N M_{\rm bar} \alpha\mu^2 \exp(-\mu R_{\rm flat}).
\end{equation}

In the empirical MOND model~\cite{Milgrom1,Milgrom2} the acceleration is given by
\begin{equation}
g_N=\nu(a/a_0)a,
\end{equation}
where $g_N$ is the Newtonian acceleration $g_N=G_NM_{\rm bar}/r^2$. We have $\nu(a/a_0)\rightarrow 1$ for $a/a_0 >> 1$ and $\nu(a/a_0)\rightarrow a/a_0$ for $a/a_0 << 1$. For the case $a/a_0 << 1$, we get for circular velocities:
\begin{equation}
\label{MONDvel}
v_{c\,{\rm MOND}}=(G_NM_{\rm bar}a_0)^{1/4}.
\end{equation}
In MOND the Tully-Fisher relation is determined as an exact empirical relation:
\begin{equation}
V_{c\,{\rm flat}\,{\rm MOND}}^4=G_NM_{\rm bar}a_{0\,{\rm MOND}}.
\end{equation}

In ref. \cite{MoffatRahvar1} the empirical baryonic Tully-Fisher relation is demonstrated to follow from the MOG predicted fits to rotational velocity data. The results are shown in Fig. 1 taken from ref.~\cite{MoffatRahvar1} for the baryonic magnitude versus ${\rm log_{10}(V_{\rm flat})}$ and for $V_{c\,{\rm flat}}^n\propto M_{\rm bar}$ the best fit yields $n=3.31$ in agreement with the data and observational scatter and errors~\cite{Verheijen}.

\begin{figure}
\centering \includegraphics[scale=0.6]{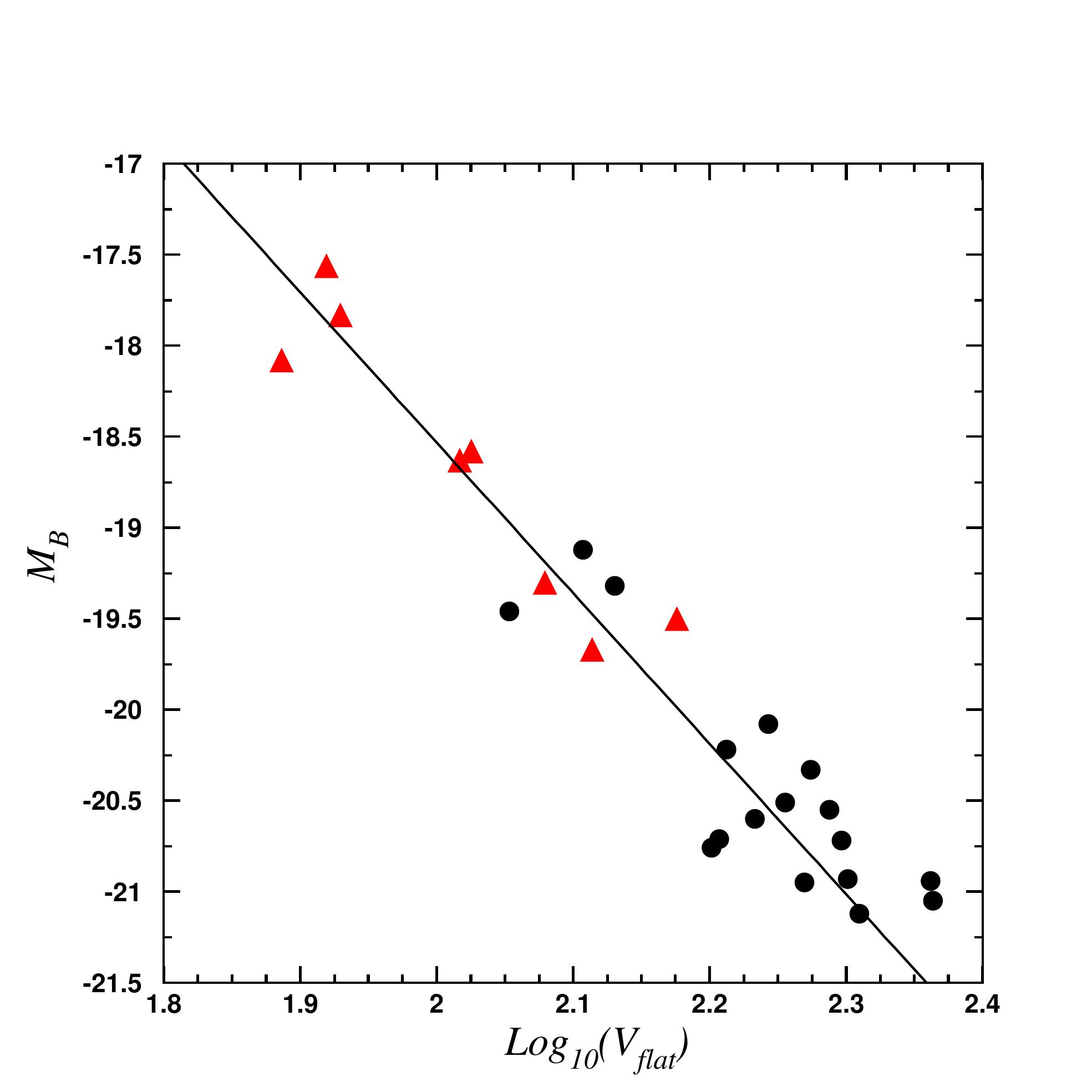}\\
\caption{Baryonic magnitude versus ${\rm log_{10}(V_{\rm flat})}$ and $V_{c\,{\rm flat}}^n\propto M_{\rm bar}$ with $n=3.31$.}{\label{fig.TullyFisher}}
\end{figure}

MOG predicts the asymptotic flatness of galaxy rotation velocities in the regions of the galaxies where stellar matter and gas are visible. However, MOG differs from the prediction of MOND for $r\rightarrow\infty$. We observe from (\ref{MONDvel}) that the rotational velocity $v_{c\,{\rm MOND}}$ remains flat and constant to infinity, while for MOG (\ref{AsymptoticMOG}) leads for asymptotically large $r$ to a more realistic Newtonian-Kepler fall-off for $v_c$. In the fits to Milky Way galaxy data~\cite{MoffatToth1}, it has been shown that MOG fits data out to $R\sim 200$ kpc, while MOND fails to fit the data for isotropic behavior of rotational velocities of stars and tracers.

\section{The Mass-Discrepancy-Baryonic Relation}

McGaugh et al.,~\cite{McGaugh} have demonstrated that the tight one-to-one correlation between $a_{\rm bar}$ and $a_{\rm obs}$ strongly suggests that ordinary baryons are the source of the gravitational potential for galaxies. The results demonstrating the correlation between the observed galaxy accelerations and the baryonic matter is described in Fig 2~\footnote[3]{The figure is reproduced with permission from ref.~\cite{McGaugh}.}. Unless a dark matter paradigm can explain this correlation, then one will be motivated to consider altering the gravitational dynamical laws rather than invoke dark matter. An empirical  functional form that provides a good fit to the data is given by~\cite{McGaugh}
\begin{equation}
\label{Empiricalacc}
a_{\rm obs}={\cal F}(a_{\rm bar})=\frac{a_{\rm bar}}{1-\exp(-\sqrt{a_{\rm bar}/a_0})},
\end{equation}
where $a_0=(1.20\pm 0.02\,({\rm random})\pm 0.24\,({\rm systematic})\times 10^{-10}\, {\rm m}\,{\rm sec}^{-2}$. The random error is a $1\sigma$ value, while the normalization of the mass-to-light ratio $\Upsilon_*$ is a $20\%$ normalization uncertainty.  Eq. (\ref{Empiricalacc}) provides a good description of $2693$ individual data points in 153 different galaxies in the SPARC sample of galaxies. Eq. (\ref{Empiricalacc}) contains the one critical acceleration scale $a_0$, and for high accelerations $a >> a_0$, it describes a linear slope and for low accelerations $a << a_0$, it gives $a \propto \sqrt{a_{\rm bar}}$. The functional behavior (\ref{Empiricalacc}) ceases to fit the data for MOG as $r\rightarrow\infty$ due to the Newtonian-Kepler asymptotic behavior of the rotational velocities $v_{c\,{\rm MOG}}$ compared to the MOND prediction for $v_{c\,{\rm MOND}}$.  However, the acceleration data is given only for visible stellar mass and gas in galaxies where MOG is in good agreement with the data. 

We derive $a_{\rm MOG}$ from (\ref{MOGaccelerationlaw}) and we have $g_{\rm bar}\equiv a_{\rm bar}=G_N{\bar M}_{\rm bar}/r^2$.  We obtain for the point particle MOG solution a fit to the McGaugh et al., galaxy data~\cite{McGaugh} adopting a mean binned mass from averaging 149 galaxy masses, ${\bar M}_{\rm bar}=6.46\times 10^{10}M_\odot$, and using the best fit values $\alpha=8.89$ and $\mu=0.042\,{\rm kpc}^{-1}$ obtained in ref.~\cite{MoffatRahvar1}. The result of this fit is shown in Fig. 3 where the full black curve is the fit to (\ref{Empiricalacc}) and the full green curve is the MOG prediction. For small accelerations the MOG prediction begins to deviate from the fit to (\ref{Empiricalacc}), a prediction that can be tested with future small acceleration data. A more general result with varying galaxy masses and radii using the SPARC data is required to fully account for the McGaugh et al., data, a result to be obtained in a forthcoming paper~\cite{GreenMoffatToth}.

\begin{figure}
\centering \includegraphics[scale=0.4] {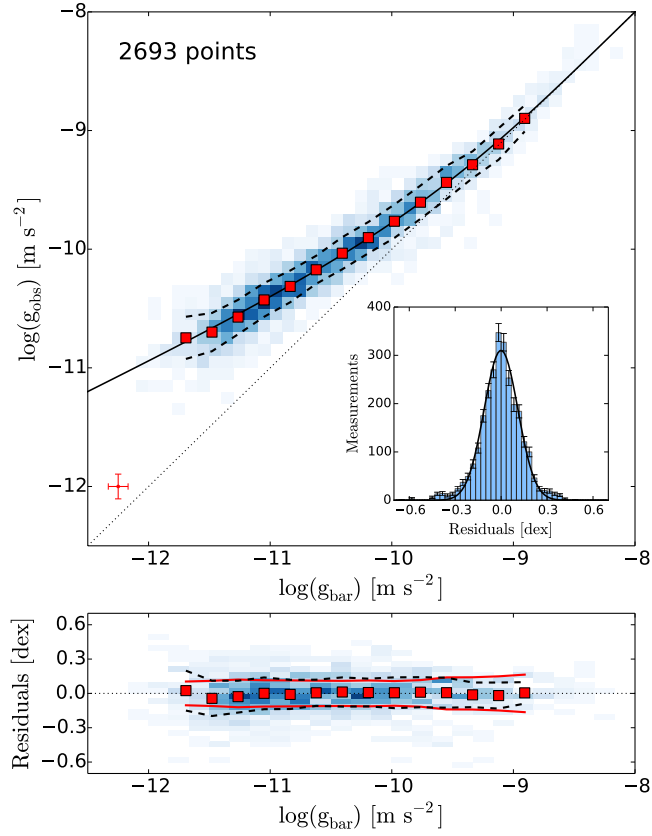}\\
\caption{Mass-discrepancy-baryonic relation~\cite{McGaugh}. The solid black curve is a fit to the empirical relation (\ref{Empiricalacc})}.{\label{fig.McGaughLelliSchombert}}
\end{figure}

\begin{figure}
\centering \includegraphics[scale=0.6]{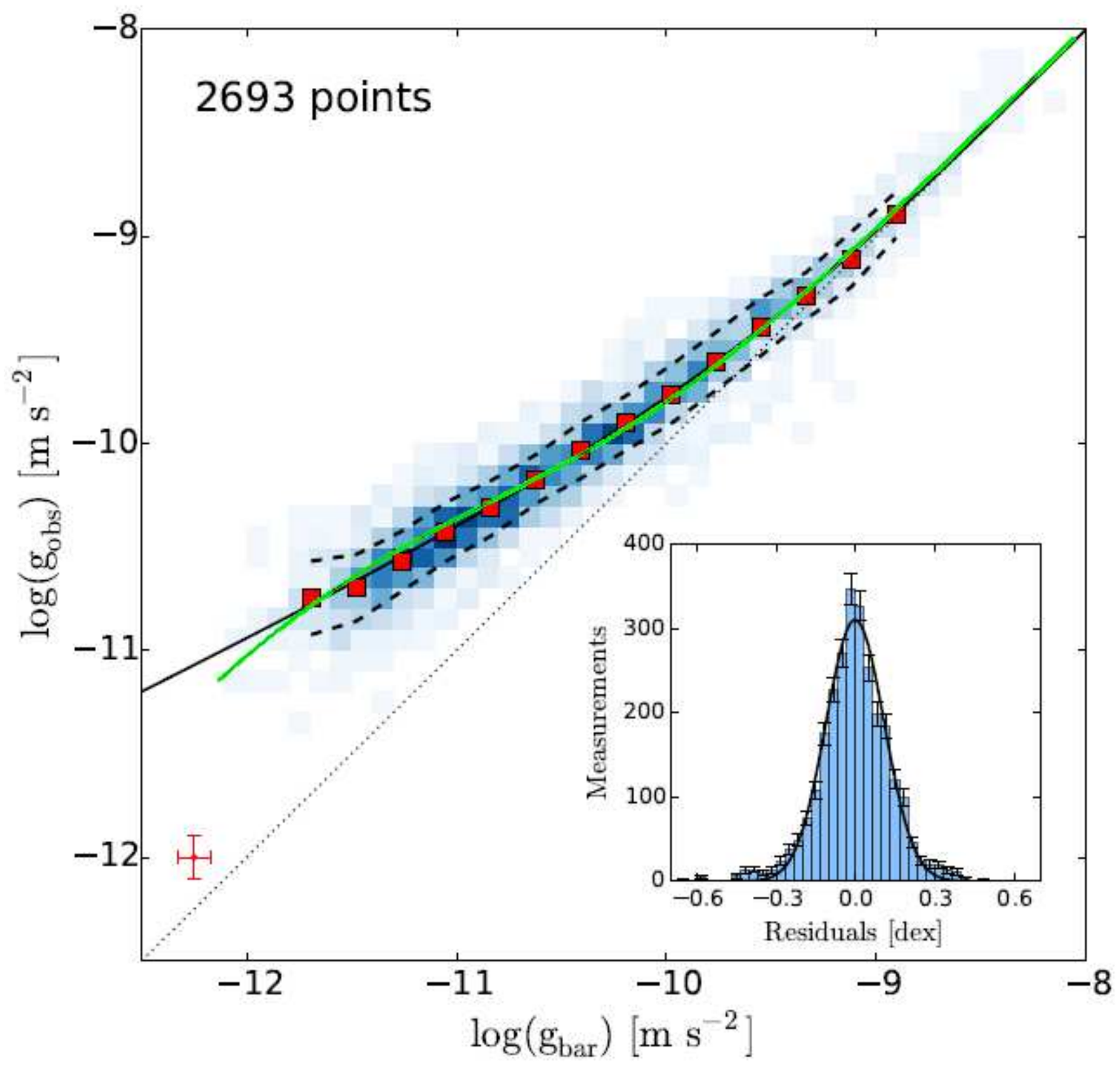}\\
\caption{The centripital acceleration $g_{\rm obs}$ is plotted against that predicted by the distribution of baryons, $g_{\rm bar}=a_{\rm bar}$. The solid black curve is a fit to the empirical formula (\ref{Empiricalacc}) and the solid green curve is the MOG prediction using $a_{\rm MOG}$.}{\label{fig.MOGMcGaughPlot}}
\end{figure}

A detailed fit of galaxies from low surface brightness to high surface brightness galaxies, including disks and bulges without dark matter, has been achieved with the best fit parameters $\alpha=8.89\pm 0.34$ and $\mu=0.042\pm 0.004\,{\rm kpc}^{-1}$~\cite{MoffatRahvar1}. A fit to dwarf galaxy data has been obtained with $\alpha=8.89$ and the average value $\mu=2.77\,{\rm kpc}^{-1}$~\cite{Haghighi}. With the choice of the best fit parameters $\alpha=8.89$ and $\mu=0.042\,{\rm kpc}^{-1}$, a good fit to cluster dynamics without dark matter has been achieved~\cite{MoffatRahvar2}. Moreover, the Bullet Cluster and the Train Wreck Cluster Abell 520 have also been fitted to the data without dark matter~\cite{IsraelMoffat}.

\section{Conclusions}

The covariant equation of motion of test bodies in the STVG version of MOG satisfies the equivalence principle, leading to the weightless free fall of bodies in a homogeneous gravitational field. The weak gravitational field and non-relativistic approximation of the MOG field equations leads to a gravitational potential and acceleration law that predict agreement with the baryonic Tully-Fisher relation and the observed flat rotation curves in galaxy data. The tight one-to-one correlation between the observed acceleration of a wide range of rotation supported galaxies and baryonic matter~\cite{McGaugh} is well fitted by MOG. The predicted Newtonian-Kepler behavior of the MOG potential and the rotation curves for large $r$ is a prediction that differs from the MOND asymptotic behavior. A first verification of the MOG asymptotic behavior as $r\rightarrow\infty$ has been obtained from a fitting of Milky Way galaxy rotation curves with a distance scale $R\sim 200$ kpc~\cite{MoffatToth1}. This prediction of the large distance scale can be further tested with future observations.

Modified gravitational theories must be tested with the widely available experimental data for lensing, merging clusters and cosmology. MOG has so-far achieved a fit to the data from the solar system to large scale cosmology~\cite{Moffat1,MoffatRahvar1,MoffatRahvar2,IsraelMoffat,MoffatToth2,MoffatRahvarToth,MoffatToth3}. 

\section{Appendix A}

The STVG field equations are given by~\cite{Moffat1}:
\begin{equation}
\label{mog1}
G_{\mu\nu}-\Lambda g_{\mu\nu}+Q_{\mu\nu}=-8\pi GT_{\mu\nu},
\end{equation}
\begin{equation}
\label{mog2}
\nabla_{\nu}B^{\mu\nu}+ \frac{\partial V(\phi)}{\partial \phi_\mu}=-J^\mu,
\end{equation}
\begin{equation}
\label{mog3}
\nabla_\sigma B_{\mu\nu}+\nabla_\nu B_{\sigma\mu}+\nabla_\sigma B_{\nu\mu}=0,
\end{equation}
\begin{equation}
\label{mog4}
\Box G=K(x),
\end{equation}
\begin{equation}
\label{mog5}
\Box\mu=L(x).
\end{equation}
We have
\begin{equation}
Q_{\mu\nu}=\frac{2}{G^2}(\nabla^\alpha G\nabla_\alpha Gg_{\mu\nu}-\nabla_\mu G\nabla_\nu G)-\frac{1}{G}(\Box Gg_{\mu\nu}-\nabla_\mu\nabla_\nu G).
\end{equation}
Moreover,
\begin{equation}
K(x)=\biggl(\frac{16\pi}{3+16\pi}\biggr)\biggl[\frac{3}{8\pi G}(1+4\pi)\nabla^\alpha G\nabla_\alpha G
- \frac{G}{2\mu^2}\Box\mu+\frac{1}{2}G^2\biggl(T+\frac{\Lambda}{4\pi G}\biggr)+\frac{1}{\sqrt{\alpha G_N}}T^{M\mu\nu}u_\nu\phi_\mu\biggr],
\end{equation}
and
\begin{equation}
L(x)=\frac{1}{G}\nabla^{\alpha}G\nabla_{\alpha}\mu+\frac{2}{\mu}\nabla^\alpha\mu\nabla_\alpha\mu+\mu^2G\frac{\partial V(\phi_\mu)}{\partial \mu}.
\end{equation}
$G_{\mu\nu}$ is the Einstein tensor $G_{\mu\nu}=R_{\mu\nu}-\frac{1}{2}g_{\mu\nu}R$, $\Lambda$ is the cosmological constant, $\Box=\nabla^\mu\nabla_\mu$, $T=g^{\mu\nu}T_{\mu\nu}$ and $G$ and $\mu$ are scalar fields.  The Ricci curvature tensor is defined by
\begin{equation}
R_{\mu\nu}=\partial_\nu{\Gamma_{\mu\sigma}}^\sigma-\partial_\sigma{\Gamma_{\mu\nu}}^\sigma+\Gamma_{\mu\sigma}^\alpha\Gamma_{\alpha\nu}^\sigma
-\Gamma_{\mu\nu}^\alpha\Gamma_{\alpha\sigma}^\sigma.
\end{equation}
The potential $V(\phi_\mu)$ for the vector field $\phi_{\mu}$ is given by\footnote[4]{The scalar field $\omega(x)$ introduced in the original STVG paper is taken to be constant and $\omega=1$.}
\begin{equation}
\label{Vphi}
V(\phi_\mu)=-\frac{1}{2}\mu^2\phi^\mu\phi_\mu.
\end{equation}

The total energy-momentum tensor is defined by
\begin{eqnarray}
T_{\mu\nu}=T^M_{\mu\nu}+T^\phi_{\mu\nu}+T^G_{\mu\nu}+T^\mu_{\mu\nu},
\end{eqnarray}
where $T^M_{\mu\nu}$ is the energy-momentum tensor for the ordinary matter, and
\begin{equation}
\label{mog7}
T^\phi_{\mu\nu}=-\frac{1}{4\pi}\biggl[B_{\mu}^{~\alpha}B_{\nu\alpha}-g_{\mu\nu}\left(\frac{1}{4}B^{\rho\alpha}B_{\rho\alpha}+V(\phi_\mu)\right)
+2\frac{\partial V(\phi_\mu)}{\partial g^{\mu\nu}}\biggr],
\end{equation}
\begin{equation}
T^G_{\mu\nu}=-\frac{1}{4\pi G^3}\biggl(\nabla_\mu G\nabla_\nu G-\frac{1}{2}g_{\mu\nu}\nabla_\alpha G\nabla^\alpha G\biggr),
\end{equation}
\begin{equation}
T^\mu_{\mu\nu}=-\frac{1}{4\pi G\mu^2}\biggl(\nabla_\mu\mu\nabla_\nu\mu-\frac{1}{2}g_{\mu\nu}\nabla_\alpha\mu\nabla^\alpha\mu\biggr).
\end{equation}

The covariant current density $J^\mu$ for matter is defined by
\begin{equation}
\label{currentdensity}
J^\mu=\kappa T^{M\mu\nu}u_\nu,
\end{equation}
where $\kappa=\sqrt{\alpha G_N}$, $\alpha=(G-G_N)/G_N$ is a dimensionless scalar field, $u^\mu=dx^\mu/ds$ and $s$ is the proper time along a particle trajectory.

\section*{Acknowledgments}

We thank Martin Green and Viktor Toth for helpful discussions. I thank Martin Green for providing Fig. 3. This research was supported in part by Perimeter Institute for Theoretical Physics. Research at Perimeter Institute is supported by the Government of Canada through the Department of Innovation, Science and Economic Development Canada and by the and by the Province of Ontario through the Ministry of Research, Innovation and Science.

\end{document}